# A comparative analysis of the effect of bobbin topography in the transmission performance capability of Hybrid corrugated plane type transmission surface for MR Clutches

Loyad Joseph Losan
*Department of Mechanical Engineering*
*National Institute of Technology*
*Calicut*
Calicut, India
loyadjosephlosan@gmail.com

Saddala Reddy Tharun
*Department of Mechanical Engineering*
*National Institute of Technology*
*Calicut*
Calicut, India
sreddytharun@gmail.com

Jithin Vijaykumar
*Department of Mechanical Engineering*
*National Institute of Technology*
*Calicut*
Calicut, India
vijayakumarjithin@gmail.com

Murthi Ram Chandra Reddy
*Department of Mechanical Engineering*
*National Institute of Technology*
*Calicut*
Calicut, India
murthiramchandrareddy@gmail.com

Mood Rahul
*Department of Mechanical Engineering*
*National Institute of Technology*
*Calicut*
Calicut, India
rc64998@gmail.com

Jagadeesha T
*Department of Mechanical Engineering*
*National Institute of Technology*
*Calicut*
Calicut, India
jagadishsg@nitc.ac.in

*Abstract*— **Magneto-Rheological (MR) fluid based devices work on the principle of changing the rheological properties of MR fluid (MRF) using magnetic field excitation generated from an electromagnet. The electromagnet is usually created with the aid of copper coils wound on a low magnetic permeable spindle structure referred to as bobbin. In this paper, an attempt has been made to investigate the different bobbin configurations and its effect on the torque transmissibility of a MR clutch (MRC). A hybrid corrugated-plane type transmission surface MRC is chosen for the study, due to the advantage of enhanced transmission capability due to the simultaneous existence of plane and corrugated extensions on the disc surface. This enhanced transmission capability, resulting from the hybrid corrugated plane type transmission surface facilitates mama,. BB vg BBC gg of the influence of bobbin configuration on the torque transmission capability in an MRC. A comparative analysis using COMSOL Multiphysics software is carried out between five different innovative bobbin configurations such as rectangular, semi-circular, conical, I-sectioned and H-sectioned. This study aims to simulate and reason the variations in the magnetic field line characteristics upon variations in bobbin topography. The results obtained testify for the need of a bobbin design taking into account the transmission surface geometry. For the specific design analyzed, it was found that the H-sectioned bobbin provided the maximum torque transmission capability when compared with other topographies, whereas the conical shaped bobbin topography proved to be least facilitating for torque transmission.**

## I. INTRODUCTION

Smart materials, also known as responsive materials, are the materials whose properties can be precisely varied reversibly by controlling the external stimuli. Magneto-rheological fluid (MRF) is one of the smart materials whose viscosity substantially depends on the density of magnetic field applied as the stimuli through current as the control variable. MRF behaves as a Newtonian fluid in the absence of an external magnetic field, but instantaneously turns into a semi-solid material, whose rheological properties can be modeled mathematically as Bingham-plastic fluid, when an electromagnet producing magnetic field is activated, and thus causes fibrillated chain formation resulting in semi-active control. These unique variable characteristics make the usage of MRF quite motivating in various domains of industrial mechanical systems such as rotating machineries like clutches and brakes, which function in the shear mode of MRF[1-5].

Xu et al. designed the MRF based clutch to regulate the cooling system fan of an engine. It was proved that the air gap present in the clutch shell has a predominant effect on the distribution of magnetic flux density in the MRF area of shear [6]. Shafer et al. developed a MR clutch for robotic applications and the prototype designed has an operating range of about 30 Hz and can transmit up to 75 Nm [7]. Bucchi et al. has explored how an MR clutch was designed to disengage auxiliary in an automobile engine. Before arriving at the final configuration, other configurations were taken into consideration using an iterative process that combined mechanical design and magnetic field analysis [8]. Saito et al. developed and evaluated a normally closed (NC) form of MR clutch, employing the MR suspension. This allowed for the implementation of a safety device capable of managing the torque output of the robot joint axis and securing the holding torque in the event of a sudden halt [9]. Kikuchi et al. presented a Leg-Robot to realise various haptic spastic movement features and three different sorts of control methods—"spasticity," "clasp-knife phenomenon," and "ankle clonus"—were offered [10]. Song et al. attempted to wind the coils directly on a magnetic bobbin which forms a part of the housing, as an alternative to winding on a separate non-magnetic bobbin. It was noted that this would increase fabrication accuracy and efficiency, lower manufacturing costs, and alleviate the problem associated with the bottlenecking problem associated with the magnetic circuit of the designed MR brakes [11].

## II. DESIGN AND MODELLING

The MR clutch comprises of the components namely:

1. Input (or driving) shaft.
2. Output (or driven) shaft.
3. Electromagnets formed by copper coils wound around a bobbin.
4. MR fluid

The topography of the hybrid corrugated plane type MR clutch studied is presented in Fig. 1. The shaft is made of structural steel (non-ferro magnetic) and the housing, along with the hybrid disc which is the combination of both plane and corrugated configurations, are made of low carbon steel which is highly magnetically permeable. The gap between the housing and the disc is occupied by the MRF.

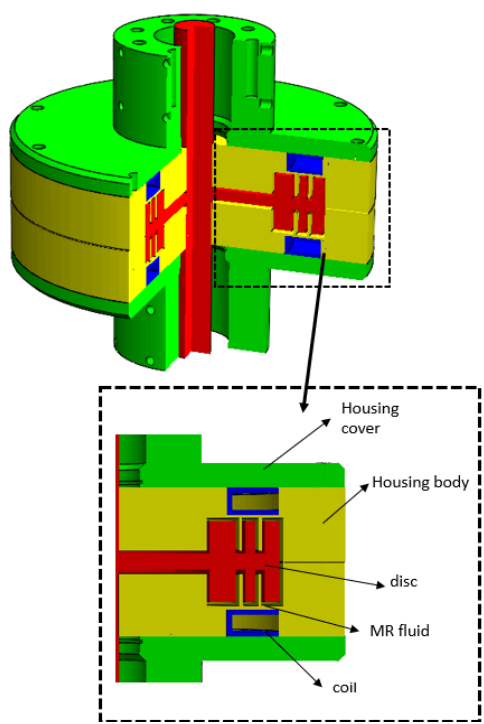

Fig. 1. Schematic design of optimized hybrid disc MR Clutch with U shaped bobbin.

Conventionally, U-shaped or rectangular bobbins are used in constraining the boundary of the electromagnetic coil. The coils are wound around a bobbin, which The present work aims to do a comparative analysis based on the magnetic field line characteristics subject to different bobbin topographies.

The maximal torque that an MR clutch can transmit depends upon the shear strength of the MR fluid which is a first hand property of the nature of MR fluid being used and the strength of the magnetic field applied. For the torque to be generated in shear mode, it is necessary that the magnetic field lines pass through the MR gap perpendicularly in order that the fibrillations of microsized particles in MRF forms a strong material chain matrix. Hence, perpendicular routing of the magnetic field lines are consequential to leveraging torque transmission capability of the MRC.

It is with regard to the proper magnetic field line routing that the bobbin material is configured to be of a comparatively low magnetically permeable material such as stainless steel. This configuration allows the magnetic field lines to be directed perpendicular to the transmission surfaces to form long loops rather than shorter ones, ultimately yielding to the necessity of the field lines passing through the MR gap.

This present study introduces a concept of varied bobbin topography so that the innovative dispersion of magnetically non-permeable material of the bobbin through various cross-sectional shapes can induce varied magnetic field line routing and thereby varying torque transmission capability. The proposed work studies the effect of five different topographies of the bobbin on the torque transmission capability of the hybrid corrugated plane type MR clutch.

The various bobbin topographies analyzed are listed below with a short description and its associated 3-D model. Also, the bobbin configuration inside the designed hybrid corrugated plane type transmission surface MR clutch is provided for reference.

The five different bobbin topographies analyzed include:

1. U-sectioned bobbin.
2. I-sectioned bobbin.
3. H-sectioned bobbin.
4. Semicircular sectioned bobbin.
5. Conical sectioned bobbin.

### A. U- sectioned bobbin

The 3-D CAD model of the proposed U-sectioned bobbin is given in Fig.2. The cross-section of the bobbin is also given alongside in a detailed view. It is seen that the cross-section is U-shaped.

The topography of a hybrid corrugated disc type MR clutch provided with U-sectioned bobbin is presented in Fig. 3. The electromagnetic copper coil is wound around the U-sectioned bobbin.

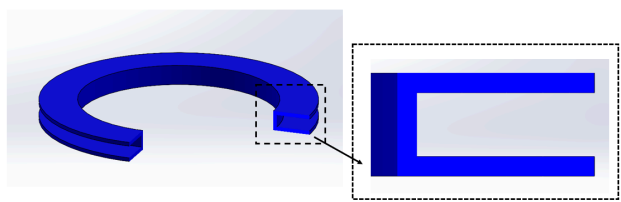

Fig.2. 3D CAD model of the U-sectioned bobbin.

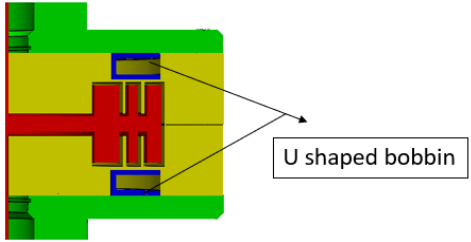

Fig. 3. The axisymmetric model of the hybrid corrugated plane type MRC with the proposed U-sectioned bobbin.

### B. I-sectioned bobbin

The 3-D CAD model of the proposed I-sectioned bobbin is given in Fig.4. The cross-section of the bobbin is also given alongside in a detailed view. It is seen that the cross-section is I-shaped.

The topography of a hybrid corrugated disc type MR clutch provided with I-sectioned bobbin is presented in Fig. 5. The electromagnetic copper coil is wound around the I-sectioned bobbin.

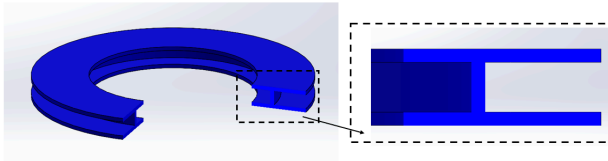

Fig.4. 3D CAD model of the I-sectioned bobbin.

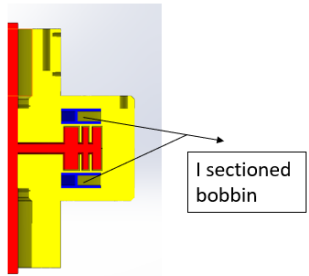

Fig.5. The axisymmetric model of the hybrid corrugated plane type MRC with the proposed I-sectioned bobbin.

### C. H-sectioned bobbin

The 3-D CAD model of the proposed H-sectioned bobbin is given in Fig.6. The cross-section of the bobbin is also given alongside in a detailed view. It is seen that the cross-section is H-shaped.

The topography of a hybrid corrugated disc type MR clutch provided with H-sectioned bobbin is presented in Fig. 7. The electromagnetic copper coil is wound around the I-sectioned bobbin on both the sides of the bobbin. This configuration in effect splits the coil into two windings, one above and one below the separation plane of the H-sectioned bobbin.

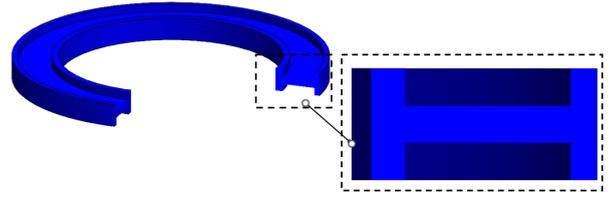

Fig.6. 3D CAD model of the H-sectioned bobbin.

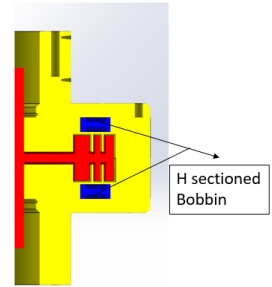

Fig.7. The axisymmetric model of the hybrid corrugated plane type MRC with the proposed H-sectioned bobbin.

### D. Semicircular sectioned bobbin

The 3-D CAD model of the proposed semi-circular sectioned bobbin is given in Fig.8. The cross-section of the bobbin is also given alongside in a detailed view. It is seen that the cross-section is semicircular in shape.

The topography of a hybrid corrugated disc type MR clutch provided with semicircular sectioned bobbin is presented in Fig.9. The electromagnetic copper coil is wound around the semi-circular sectioned bobbin.

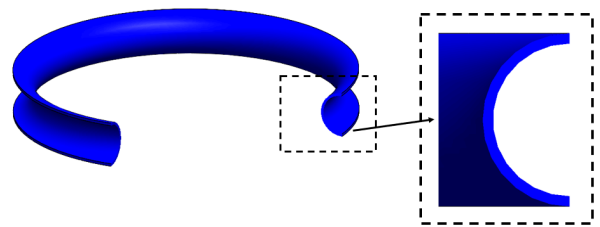

Fig.8. 3D CAD model of the semicircular sectioned bobbin.

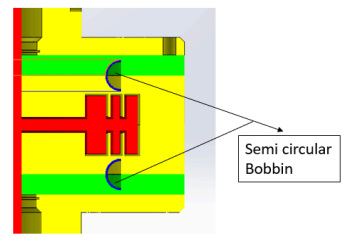

Fig.9. The axisymmetric model of the hybrid corrugated plane type MRC with the proposed semicircular sectioned bobbin.

## E. Conical sectioned bobbin.

The 3-D CAD model of the proposed Conical sectioned bobbin is given in Fig.10. The cross-section of the bobbin is also given alongside in a detailed view. It is seen that the cross-section is Conical in shape.

The topography of a hybrid corrugated disc type MR clutch provided with conical sectioned bobbin is presented in Fig.11. The electromagnetic copper coil is wound around the conical sectioned bobbin.

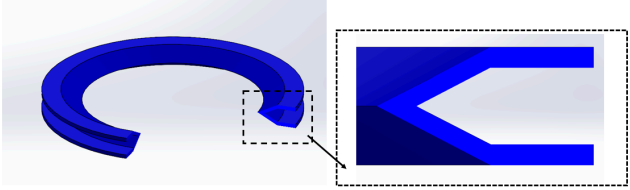

Fig.10. 3D CAD model of the conical sectioned bobbin.

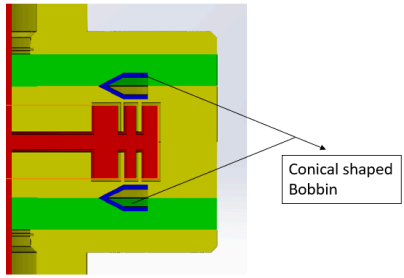

Conical shaped Bobbin

Fig.11. The axisymmetric model of the hybrid corrugated plane type MRC with the proposed conical sectioned bobbin.

### III. MATHEMATICAL MODELING

The torque transmitted by the MR Clutch upon actuation by the current can be determined by using the basic torque equation given in equation 2.

$$dT = 2\pi r^2 \tau_z dr + 2\pi r^2 \tau_r dz \quad (2)$$

where r is the radial distance from the axis, $\tau_z$ and $\tau_r$ are the shear stresses along with axial and radial directions. The shear stress behavior of the MR fluid can be modeled using the Bingham-plastic model.

The term $2\pi r^2 \tau_z dr$ give the torque from a radial surface while the term $2\pi r^2 \tau_r dz$ gives the torque from an axial surface.

$T_1$, the torque contributed by a single radial surface is given by:

$$T_1 = 2\pi \int_{r_1}^{r_2} r^2 \tau_z dr \quad (3)$$

$T_2$, the torque contributed by a single axial surface is given by:

$$T_2 = 2\pi r^2 \int_{z_1}^{z_2} \tau_r dz \quad (4)$$

The total output torque of the clutch can be found by adding together all the individual torques from all the radial and axial surfaces.

$$T = \sum_{i=1}^{n} T_1 + \sum_{j=1}^{m} T_2$$

where, $n$ is the number of radial surfaces and $m$ is the number of axial surfaces in the axisymmetric model of the MR Clutch.

### IV. MAGNETIC ANALYSIS

The magnetic analysis for the five different topographical configurations of the MRC was performed using COMSOL Multiphysics software. The material was used for all the various parts of the MR clutch like the housing, the disc, the coil are listed in TABLE I. Also, the parameters listed in TABLE II are kept constant throughout the magnetic analysis for the five different bobbin configurations.

Maxwell's equations were solved by the non-linear solver using finite element techniques, and using the built-in mathematical tools, the torque transmission capability was evaluated. A 2-D axi-symmetric analysis with Triangular elements was employed in order to discretize the complex geometry and hence execute the computation.

It is of interest to the present discussion to note that for each of the five magnetic simulations conducted for the different bobbin configurations, the number of coils were kept constant so as to infer an unbiased effect of the bobbin geometry. For each of the bobbin configurations, the area of the coil configuration inside the bobbins was kept constant and the parameters associated to that geometry varied to achieve this.

TABLE I. MATERIALS USED AND MAGNETIC PROPERTIES OF VARIOUS COMPONENTS

| Component | Material | Relative Permeability | Saturation Flux Density |
|---|---|---|---|
| Magnetic core | Low Carbon Steel | B-H Curve | 1.55 T |
| Coil | Copper | 1 | - |
| MR fluid | AMT Magnaflow + | B-H Curve | 1.65 T |
| Bobbin and others | Structural Steel | 1 | - |

| Geometrical parameters of the MRC | Value used for simulation |
|---|---|
| Clutch radius | 78 mm |
| Clutch width | 60 mm |
| Number of turns of coil per bobbin | 300 |
| Current in the coil | 2.5 A |
| bobbin_width | 1.6 mm |
| Width of disc | 4.98 mm |
| Height of corrugation 1 | 10 mm |
| Height of corrugation 2 | 10 mm |
| Height of corrugation 3 | 10 mm |
| Radial extent of the plane transmission surface | 36.5 mm |

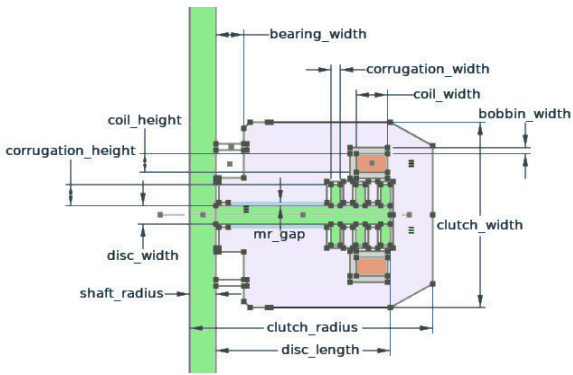

Fig. 12. Schematic diagram of cross-section of hybrid corrugated disc type MR clutch.

## A. U-sectioned bobbin

The magnetic analysis for the MRC design with a U-sectioned bobbin, given in Fig.14 revealed that the torque transmission capability of the hybrid corrugated disc clutch, is 130.03 Nm, for the specified design with design parameters listed in Table II . Fig.14 depicts the path of the magnetic flux lines forming closed loops. The analysis results help us in comprehending the efficiency of the bobbin topography through the inference of magnetic field lines. Fig.13 depicts the geometrical parameters of the U-sectioned bobbin. Table III lists the numerical values chosen for the geometrical parameters of the U-sectioned bobbin referenced in Fig.13.

The U-sectioned bobbin and hence the coil have allowed for the majority of the field lines to pass perpendicular to the annular MR gap. The coil acquired the geometry from the bobbin cross section. This is reflected in the directionality of the magnetic field lines which in this case replicates a U-shaped loop. This ensures that the annular portion of the disc contributes significantly to the torque transmission of the MRC, as the field lines can be observed to be perpendicular to the transmission surface of the MRC.

TABLE III. GEOMETRIC PARAMETERS OF THE U-SECTIONED BOBBIN

| Geometrical parameters of the bobbin | Value |
|---|---|
| bobbin_width | 0.0016087 m |
| coil_height | 0.0044051 m |
| coil_width | 0.013968 m |
| radial position of the bobbin | 0.051174 m |

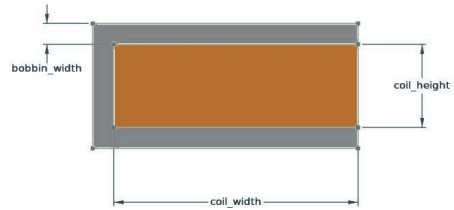

Fig. 13. Schematic diagram of the cross-section of the U shaped bobbin.

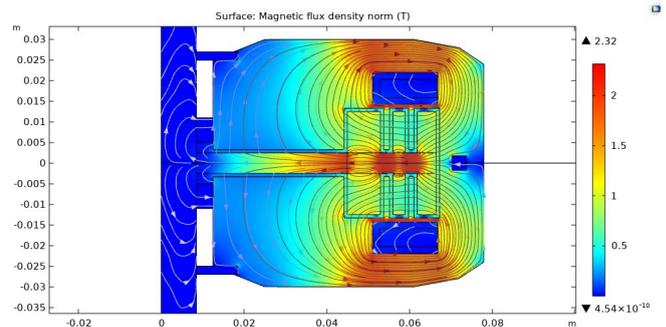

Fig. 14. Surface plot of the magnetic field density when a U sectioned bobbin is configured in the MRC design.

## B. I-sectioned bobbin

The magnetic analysis for the MRC design with an I-sectioned bobbin, given in Fig.16 revealed that the torque transmission capability of the hybrid corrugated disc clutch is 113.37 Nm, for the specified design with parameters listed in Table II . Fig.16 depicts the path of the magnetic flux lines forming closed loops. Fig.15 depicts the geometrical parameters of the I-sectioned bobbin. Table IV lists the numerical values chosen for the geometrical parameters of the I-sectioned bobbin referenced in Fig.15.

The I-sectioned bobbin and hence the coil have allowed for the majority of the field lines to pass perpendicular to the annular MR gap, even though the routing was not as efficient as was seen for in the case of U-sectioned bobbin of Fig.14. Also, the flimsiness of the bobbin with a thickness of 1.6 mm can be remedied by the provision of the I-section thereby increasing the structural integrity of the part from the light of manufacturing ease and service care. The coil acquires the geometry from the bobbin cross section. This is reflected in the directionality of the magnetic field lines which in this case replicates a U-shaped loop, but with the support length of the I-section prolonging the loop formation of the magnetic field lines. This ensures that the annular as well as radial portions of the disc contributes to the torque transmission of the MRC, as the majority of field lines can be observed to be perpendicular to the transmission surface of the MRC. The analysis has been done with air being assigned as the material between the inner support lengths of the I-sectioned bobbin.

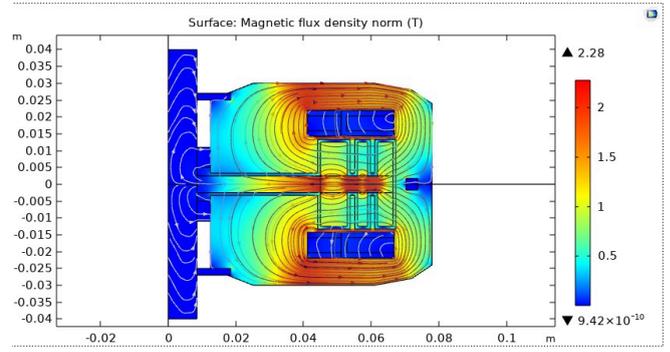

Fig. 16. Surface plot of the magnetic field density when a I sectioned bobbin is configured in the MRC design.

### C. H sectioned bobbin

The magnetic analysis for the MRC design with a H-sectioned bobbin, given in Fig.18 revealed that the torque transmission capability of the hybrid corrugated disc clutch, is 137.6 Nm, for the specified design with parameters listed in Table II . Fig.18 depicts the path of the magnetic flux lines forming closed loops. Fig.17 depicts the geometrical parameters of the H-sectioned bobbin. Table V lists the numerical values chosen for the geometrical parameters of the H-sectioned bobbin referenced in Fig.17.

The H-sectioned bobbin has allowed for more of the field lines to pass perpendicular to the annular as well as radial  MR gaps compared to the U-sectioned bobbin. The H sectioned bobbin serves as a way to keep the number of turns of the coil constant while increasing the width of the coil and not decreasing the height of the coil. A decrease in the height of the coil will decrease the dispersion of magnetic field lines and hence the magnetic flux density in the MR domain. This configuration serves as a way to split the coil into multiple components so as to increase the area cast by the field lines. The coil acquires the geometry from the bobbin cross section and the configuration of two sets of coils allows the field lines to be directed further deep into the inner radii of the clutch, thereby resulting in the provision of 137.6 Nm torque transmission capability. Also, the structural integrity resulting from a H-section provides sturdiness as well as margin for  compromise on care during manufacturing, thereby reducing the cost.

TABLE IV.          GEOMETRIC PARAMETERS OF THE I SECTIONED BOBBIN

| Geometrical parameters of the bobbin | Value |
|---|---|
| Bobbin thickness | 0.0016087 m |
| coil_height: | 0.0044051  m |
| coil_width | 0.013968 m |
| I_section_width | 10 mm |
| radial position of the bobbin | 0.0667507 m |

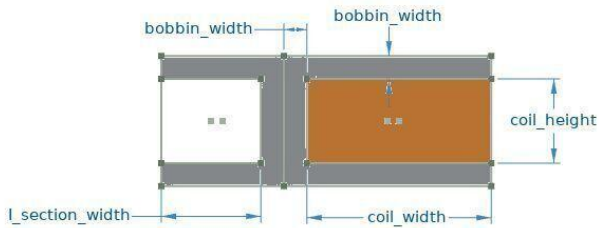

Fig. 15. Schematic diagram of the cross-section of the I shaped bobbin.

TABLE V. GEOMETRIC PARAMETERS OF THE H SECTIONED BOBBIN

| Geometrical parameters of the bobbin | Value |
|---|---|
| bobbin_width | 0.0016087 m |
| coil_height: | 0.012 m |
| coil_width | 0.00513 m |
| radial position of the bobbin | 0.0475 m |

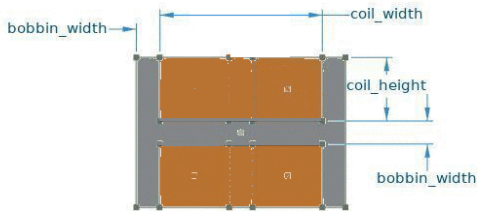

Fig. 17. Schematic diagram of the cross-section of the H shaped bobbin.

| Geometrical parameters of the bobbin | Value |
|---|---|
| bobbin_thickness | 0.0016087 m |
| coil_radius | 0.00626 m |
| radial position of the bobbin | 0.0590427 m |

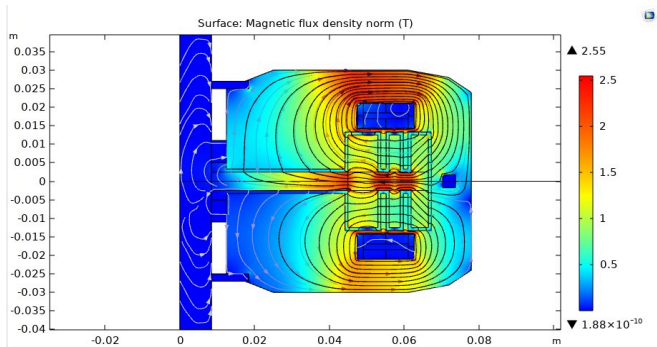

Fig. 18. Surface plot of the magnetic field density when a H sectioned bobbin is configured in the MRC design.

*D. Semicircular bobbin*

The magnetic analysis for the MRC design with a semicircular-sectioned bobbin, given in Fig.20 revealed that the torque transmission capability of the hybrid corrugated disc clutch was only 3.4615 Nm, for the specified design with parameters listed in Table II . Fig.20 depicts the path of the magnetic flux lines forming closed loops in the MRC system. Fig.19 depicts the geometrical parameters of the semicircular-sectioned bobbin. Table VI lists the numerical values chosen for the geometrical parameters of the semicircular-sectioned bobbin referenced in Fig.19.

In the present design configuration of the MRC a semi circular sectioned bobbin was found to be the least efficient. The width of such a cross section is half that of the height ,the decrease in width is having a major impact on the routing of the magnetic flux lines, thus failing to pass perpendicular to the annular cross section that contributes more to the transmission in the case of other bobbin designs. The height of the bobbin being large also impacts the proper flux path negatively. Since the bobbin is circular near the annular gaps ,most of the field lines do not reach the annular MR gap perpendicular.

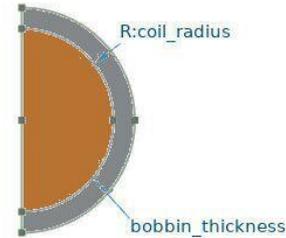

Fig. 19. Schematic diagram of the cross-section of the semi circle shaped bobbin.

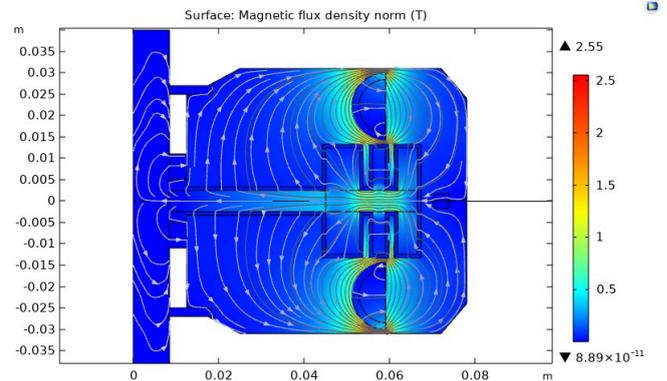

Fig. 20. Surface plot of the magnetic field density when a semi circular sectioned bobbin is configured in the MRC design.

*E. Conical sectioned bobbin*

The magnetic analysis for the MRC design with a conical sectioned bobbin, given in Fig.22, revealed that the torque transmission capability of the hybrid corrugated disc clutch, is 114.03 Nm, for the specified design with parameters listed in Table II . Fig.22 depicts the path of the magnetic flux lines forming closed loops. Fig.21 depicts the geometrical parameters of the conical-sectioned bobbin. Table VII lists the numerical values chosen for the geometrical parameters of the conical-sectioned bobbin referenced in Fig.21.

The conical-sectioned bobbin and its geometry as acquired by the coil have allowed for the majority of the field lines to pass perpendicular to the annular MR gap.

Also, due to the mix of U-sectioned as well as the triangular curved structure of the conical shaped bobbin, the field lines directionality is able to exploit both the radial as well as the annular transmission surfaces. The coil acquires the geometry from the bobbin cross section. This is reflected in the directionality of the magnetic field lines which in this case replicates a hybrid U-shaped loop near the rectangular region of the bobbin and a triangular field line routing near to the triangular portion of the bobbin. This ensures that both the radial as well as the annular portion of the transmission surfaces contributes to the torque capability of the MRC.

TABLE VII.    GEOMETRIC PARAMETERS OF THE CONICAL SECTIONED BOBBIN

| Geometrical parameters of the bobbin | Value |
|---|---|
| bobbin_width | 0.0016087 m |
| coil_height | 0.006 m |
| coil_width | 0.008 m |
| cone_angle | 70 degree |
| radial position of the bobbin | 0.0475 m |

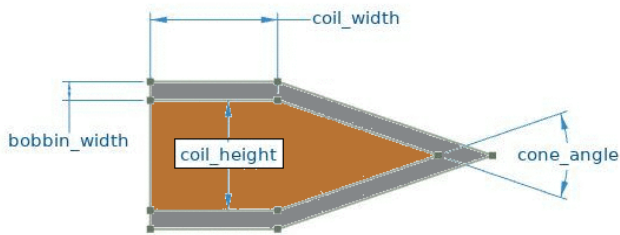

Fig. 21. Schematic diagram of the cross-section of the cone shaped bobbin.

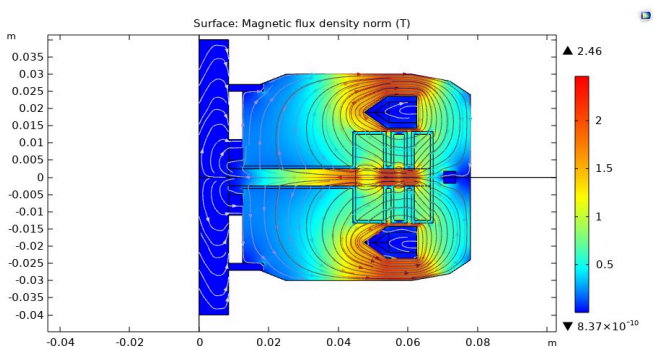

Fig. 22. Surface plot of the magnetic field density when a conical sectioned bobbin is configured in the MRC design.

V.    RESULTS AND DISCUSSIONS

The present analytical study dealt with the torque transmission capability for five different bobbin topographies. It was observed that for the same design parameters of influence, the torque transmission capability was highest for the H-sectioned bobbin topography where the concept of coil splitting was employed. This facilitated the field line to be routed over a larger curvature so as to utilize the available transmission surfaces.

The least performer out of all the bobbin topographies was the semicircular bobbin configuration due to the poor routing of magnetic field lines perpendicular to the transmission surfaces. The descending order of the bobbin topographies with respect to the transmission capability for the hybrid corrugated plane type transmission surfaces are H-sectioned bobbin, U-sectioned bobbin, Conical-sectioned bobbin, I-sectioned bobbin and the semi-circular bobbin.

TABLE VIII.    COMPARISON OF THE RESULTS FOR FIVE DIFFERENT BOBBIN TOPOGRAPHIES

| Bobbin Topography | Achieved Torque Transmission capability |
|---|---|
| U-sectioned bobbin | 130.03 Nm |
| I-sectioned bobbin | 113.37 Nm |
| H sectioned bobbin | 137.6 Nm |
| Semicircular bobbin | 3.46 Nm |
| Conical sectioned bobbin | 114.03 Nm |

VI.    CONCLUSION

The present study examined the effect of bobbin topography on the torque transmission capability of hybrid corrugated plane MRC. It is proved from the varying torque transmission capability for the bobbin topographies that the magnetic field lines and their directionality can be controlled by the use of a properly architectured bobbin. The field lines inherit the geometry of the coil geometry, which in turn is dependent upon the bobbin topography. This topographical design allows for efficient as well as alterable transmission capability in MR clutches.

Also, it was observed that the splitting up of coils into multiple components without altering the total number of turns actually augmented the torque transmission capability for the hybrid corrugated plane type MR Clutches.

Lastly, the important takeaway from the present study is the requirement of a topographical design of bobbin taking into account the configuration of the transmission surfaces.